\documentclass[aps,amsmath,amssymb,twocolumn,superscriptaddress,nofootinbib]{revtex4-1}

\usepackage{graphicx}
\usepackage{dcolumn}
\usepackage{bm}
\usepackage{color}
\usepackage[colorlinks=true,pdfstartview=FitV,linkcolor=blue,citecolor=blue,urlcolor=blue]{hyperref}
\usepackage[mathlines]{lineno}
\usepackage[dvipsnames]{xcolor}
\usepackage[utf8]{inputenc}
\usepackage[T1]{fontenc}
\usepackage{newtxtext,newtxmath}
\usepackage[dvipsnames]{xcolor}
\usepackage{multirow}
\usepackage{threeparttable}
\usepackage{booktabs}
\usepackage{float}
\usepackage{amsmath}
\usepackage{gensymb}
\everymath{\displaystyle}

\usepackage{enumitem}
\setlist[enumerate]{label*=\arabic*.}

\usepackage{xcolor}

\newcommand{\powert}[2]{\mbox{${#1}^{#2}$}}

\newcommand{\muevc}{\mbox{$\mu$eVc$^{-2}$}}
\newcommand{\evc}{\mbox{eVc$^{-2}$}}

\newcommand{\km}{\mbox{$\chi$}}

\newcommand{\hpfreq}{\mbox{$\nu_{\rm HP}$}}
\newcommand{\mdm}{\mbox{$m_{\rm HP}$}}

\begin{document}

\title{Wideband Direct Detection Constraints on Hidden Photon Dark Matter \\ with the QUALIPHIDE Experiment}

\author{K.~Ramanathan}\email{karthikr@caltech.edu}\affiliation{Division of Physics, Mathematics and Astronomy, California Institute of Technology, Pasadena, CA 91125, USA} 
\author{N.~Klimovich}\email{klimovich@caltech.edu}\affiliation{Division of Physics, Mathematics and Astronomy, California Institute of Technology, Pasadena, CA 91125, USA} 
\author{R.~Basu Thakur}\affiliation{Division of Physics, Mathematics and Astronomy, California Institute of Technology, Pasadena, CA 91125, USA} 
\author{B.~H.~Eom}\affiliation{Jet Propulsion Laboratory, California Institute of Technology, Pasadena, CA 91109, USA} 
\author{H.~G.~Leduc}\affiliation{Jet Propulsion Laboratory, California Institute of Technology, Pasadena, CA 91109, USA}
\author{S.~Shu}\affiliation{Division of Physics, Mathematics and Astronomy, California Institute of Technology, Pasadena, CA 91125, USA}
\author{A.~D.~Beyer}\affiliation{Jet Propulsion Laboratory, California Institute of Technology, Pasadena, CA 91109, USA}
\author{P.~K.~Day}\affiliation{Jet Propulsion Laboratory, California Institute of Technology, Pasadena, CA 91109, USA}

\date{\today}

\begin{abstract}
We report direction detection constraints on the presence of hidden photon dark matter with masses between 20\textendash30~\muevc, using a cryogenic emitter-receiver-amplifier spectroscopy setup designed as the first iteration of QUALIPHIDE (QUAntum LImited PHotons In the Dark Experiment). A metallic dish sources conversion photons, from hidden photon kinetic mixing, onto a horn antenna which is coupled to a C band kinetic inductance traveling wave parametric amplifier, providing for near quantum-limited noise performance. We demonstrate a first probing of the kinetic mixing parameter \km\ to just above \powert{10}{-12} for the majority of hidden photon masses in this region. These results not only represent stringent constraints on new dark matter parameter space, but are also the first demonstrated use of \textit{wideband} quantum-limited amplification for astroparticle applications.
\end{abstract}

\maketitle


The nature of dark matter (DM) remains an open question, with recent astroparticle community reports stressing the need for experiments to look for sub-\evc\ mass wave-like candidates \citep{{battaglieri2017us,*kolb2018basic}}. Hidden photons, hypothesized massive vector bosons that mix with the ordinary photon \cite{{okun1982,*holdom1985}}, are compelling dark matter candidates that fit this mold. Their interaction strength with the standard model is set by the kinetic mixing \km\ between that of a "hidden" field strength tensor and regular electromagnetism (EM). Theoretical work over the last decade has explored how these hidden photons can be produced in the early universe through processes such as the misalignment mechanism and act as either the entirety of dark matter or merely a component within it that weakly couples to the standard model \cite{{Arias_2012, *Nelson_2011,*redondo2009massive,*FILIPPI2020100042}}. Crucially, this mixing sources a global oscillating ordinary electric field in free-space with an average amplitude of $\sqrt{\langle|E_\text{HP}|^2\rangle} = \chi \sqrt{2\rho_\text{DM}}$ and frequency $\nu_\text{HP} \approx  0.24 \text{ GHz} \left(m_{\rm DM}/\mu\text{eV} \right)$ \cite{{horns2013searching,*jaeckel2013antenna, *jaeckel2016directional}} for hidden photon dark matter mass $m_{\rm DM}$ and local dark matter density $\rho_{\rm DM}$.  

Any conducting surface with EM boundary condition $E_{||} = 0$, such as a metallic plate ("dish"), will source radiation perpendicular to its surface due to hidden photon conversion effects based on the existence of E$_{\rm HP}$ \cite{horns2013searching}. Due to geometrical effects, a spherical cap dish of area $A_{\rm d}$ will concentrate an emission power $P \propto \chi^2 \rho_\text{DM} A_\text{d}$ onto the radial center of the sphere. Coupling this radiation at frequency \hpfreq\ into a receiver setup allows one to spectroscopically search for the signature of any extant hidden photon dark matter through a "dish haloscope" experiment. For a model of hidden photons as a gas with random orientations of velocity relative to the surface, coupled to a single polarization antenna at the focal point, and a standard literature assumption of $\rho_{\rm DM}=0.3$~GeV$\cdot$cm$^{-3}$, the expected sensitivity for \km\ given a minimum detectable power $P_\text{det}$ will be,
\begin{equation}
    \chi_\text{sens} = 9\times 10^{-14} \left(\frac{P_\text{det}}{10^{-23} \text{W}} \right)^{\frac{1}{2}}  \left(\frac{1 \text{m}^2}{A_\text{d}} \right)^{\frac{1}{2}} \epsilon_\text{c}^{-\frac{1}{2}} 
    \label{eq:DMpower}
\end{equation}
where we have introduced a coupling efficiency $\epsilon_c$ to account for signal degradation due to geometric mismatches and transmission losses. In the language of radiometry, the Signal-to-Noise ratio of such a measurement can be expressed as,
\begin{equation}
    {\rm SNR} = \frac{P_{\rm det}}{k_{\rm B}T_{\rm sys}}\sqrt{\tau/\Delta}
\end{equation} for measurement bandwidth $\Delta$ and exposure time $\tau$.

\begin{figure*}[!htp]
    \centering
    \includegraphics[width=\textwidth]{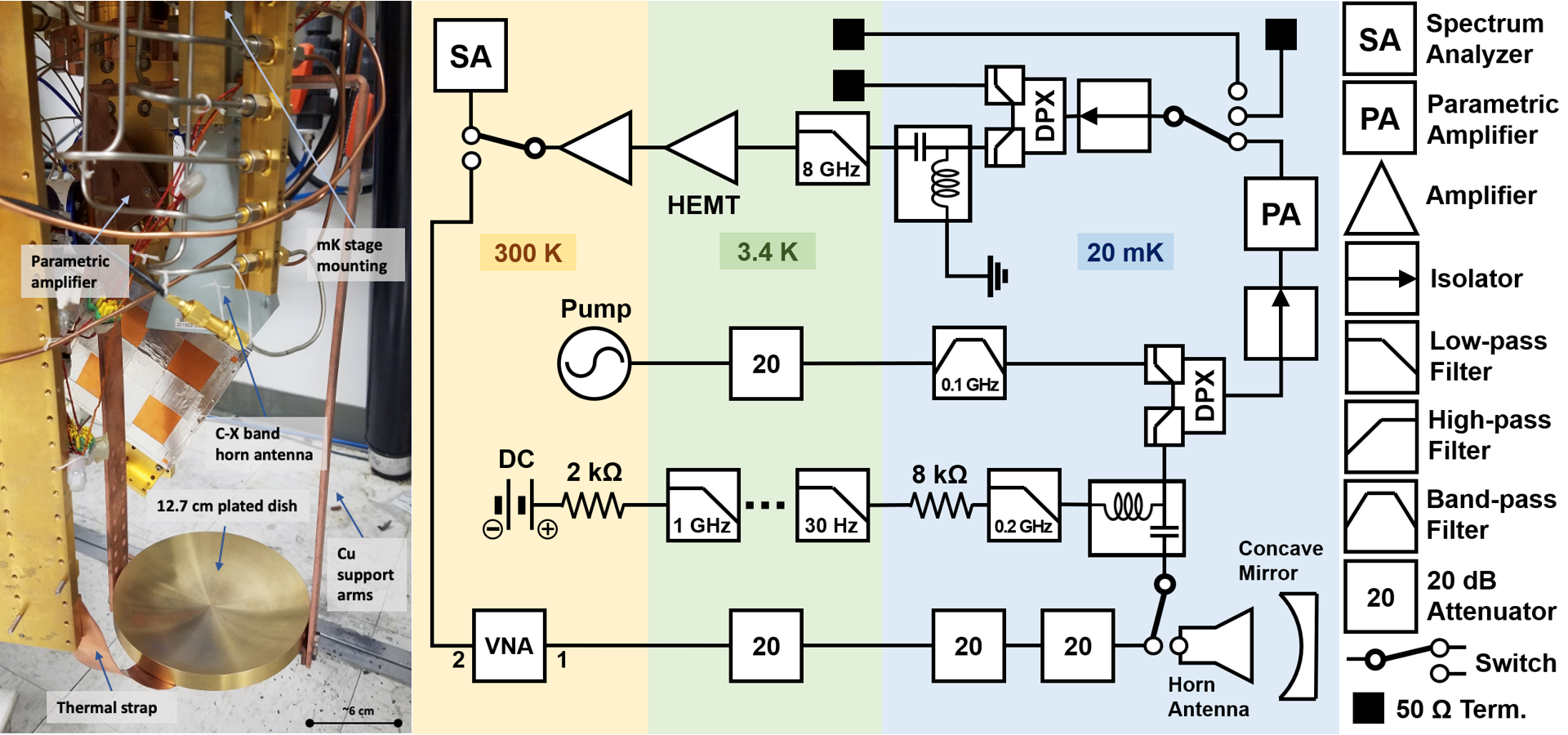}
    \caption{Left: Photograph of the QUALIPHIDE setup showing the mK stage dish, support, and antenna components. Right: cryogenic-to-room temperature schematic of the experimental layout, as described in text.}
    \label{fig:setup}
\end{figure*}

This method of dark matter detection offers an alternative to more traditional cavity searches such as ADMX or others \cite{ADMX, Jaeckel_2008} and are a hotly pursued next-generation detector scheme for axion \& axion-like particles \cite{{egge2020first,*liu2022broadband}}. The main advantage of dish searches stems from their potentially large instantaneous bandwidth which is only limited by the coupling of the antenna and readout electronics. Such experiments can provide comparable sensitivity to cavity searches with quality factor $Q$ if the dish area is sufficiently large as compared to the dark matter wavelength $A_\text{d} \sim Q\lambda_\text{DM}^2$. 

Given fixed measurement parameters then, one can improve experimental sensitivity by increasing $P_{\rm det}$ \textemdash\ most easily by increasing the emission area, an approach taken by experiments such as \cite{SHUKET,TokyoDM,KbandDM1,FUNK, knirck2018first,egge2020first} with large dishes of $A_{\rm d}>$1~m$^2$ kept at room temperature. Despite this progress, a large amount of \km-\mdm\ parameter space remains unexplored, particularly at \hpfreq\ $>$7~GHz. These frequencies are notable due to the relatively weak limits set by Cosmic Microwave Background \cite{Mirizzi_2009} and stellar measurements \cite{An_2013}. In this paper, we take the alternative approach of significantly reducing the system noise temperature $T_{\rm sys}$ with the first iteration of QUALIPHIDE (QUAntum LImited Photons In the Dark Experiment), an experiment extending the work started by \cite{KbandDM1} to its logical limit by placing a dish haloscope setup into a 20~mK environment and coupling the output to a quantum-limited travelling-wave kinetic inductance parametric amplifier (TW-KIPA). 

TW-KIPAs are state of the art devices with demonstrated 20~dB gain over an octave or more of bandwidth at GHz frequencies, with design and operation details as laid out in \cite{{klimovich2022traveling,*ho2012wideband}}. Briefly, the amplifier is able to take an input pump tone of frequency $\nu_{\rm p}$ and transfer power to an input signal tone of frequency $\nu_{\rm s}$, subsequently generating an amplified output signal along with an "idler" tone at $\nu_{\rm i}=\nu_{\rm p}-\nu_{\rm s}$, with a signal gain of approx. $G_{\rm pa} \propto {\rm exp}(k_p|I_p|^2 / (4(I_*^2 + I_{\rm DC}^2)))$ where $I_p$ is the pump current, $I_*$ is a material and geometry dependent parameter that sets the scale of the nonlinearity in the kinetic inductance, and $I_{\rm DC}$ is an applied DC current. Crucially, the end result is a signal tone with a `quantum limited' output noise $N_{\rm pa}$ of at least one quanta ($\equiv$~h$\nu$)\footnote{$N_{\rm pa}\geq \frac{1}{2} + \frac{1}{2}\left(1 - \frac{1}{G_{\rm pa}} \right)$, vacuum fluctuation and amplifier added noise respectively, as referred to the input of the amplifier.} \cite{caves1982quantum}. Going below this floor requires exotic quantum techniques like quadrature squeezing, experimentally demonstrated to only provide about 4 dB of variance reduction at present \cite{{backes2021quantum, *brubaker2017haystac}}, to make further noise reduction progress.

Fig. \ref{fig:setup} shows a labeled diagram (\textit{left}) and schematic (\textit{right}) of the QUALIPHIDE setup. A 12.7~cm diameter gold-plated copper dish ($A_{\rm d}=0.0127$~m$^2$) manufactured to $\mathcal{O}$(mm) precision, chosen for its very low thermal emissivity of $\varepsilon_{\rm s}\approx 2\times10^{-3}$ \cite{bock1995emissivity}, is attached using a copper frame and pointed to a 4.75\textendash11~GHz commercial horn antenna (AINFO LB-475110-10-C-SF) placed 22~cm away at the antenna phase center. The entire structure is mounted to the final mK stage of a Leiden dilution refrigerator. We calculate the background radiation equivalent NEP for the setup to be $2\times10^{-23}$~W/$\sqrt{\rm Hz}$ \cite{benford1998noise}. The antenna is routed to the TW-KIPA, with associated circuitry for introducing and dumping the pump tone and DC current, after which it is amplified by a commercial HEMT amplifier (Low Noise Factory LNC0.3\_14B, $T_{\rm noise}\sim3$~K) and further room temperature amplification, before being sent to a Signal Hound SM200B spectrum analyzer. A cold switch before the input of the parametric amplifier allows for the incident signal to be switched between the antenna-dish setup and 20~mK load used for reference data. A second cold switch immediately following the parametric amplifier connects to both a "hot" 3.38~K and "cold" 20~mK load used for calibrating the system noise.

The dish area, choice of antenna location, and single run configuration were limited by available space and time in the refrigerator as the measurement was conducted parasitically to existing experiments. The signal at this frequency and configuration is expected to form a convergent spherical wave with a circular focal spot \cite{horns2013searching,SHUKET}. However, due to expected diffraction effects among others, we introduce a coupling efficiency $\epsilon_{\rm c}$ consisting of three multiplicative coefficients to fully model the frequency dependent antenna-dish power coupling. The first component, between 0.79\textendash0.84 in the frequency regime of interest, accounts for imperfect power coupling and is numerically computed from an overlap integral between the focal pattern and the antenna gain pattern taken from the manufacturer datasheet, using a Gaussian beam analysis similar to \cite{knirck2018first, Goldsmith1997}. Next, $\epsilon_{\rm c}$ encapsulates a conservative attenuation term of 0.39 (-4 dB) to account for signal degradation between the antenna switch and the parametric amplifier due primarily to the placement of the first isolator. Finally, we measure the angular deviation of the dish-antenna setup and estimate its alignment as within 2$^{\degree}$ of vertical, which introduces a power coupling systematic into $\epsilon_c$ of 5\%. We use a simplified COMSOL 5.5 simulation consisting of the dish and antenna setup to validate the constrained geometry and verify the expected signal coupling to within 20\% across the frequency range. Angular shifts of the signal due to dark matter velocity dispersion effects are also considered but are calculated to introduce only negligible deviations of $\ll$1\% \cite{jaeckel2016directional}.

\begin{figure}[!htp]
    \centering
    \includegraphics[width=0.45\textwidth]{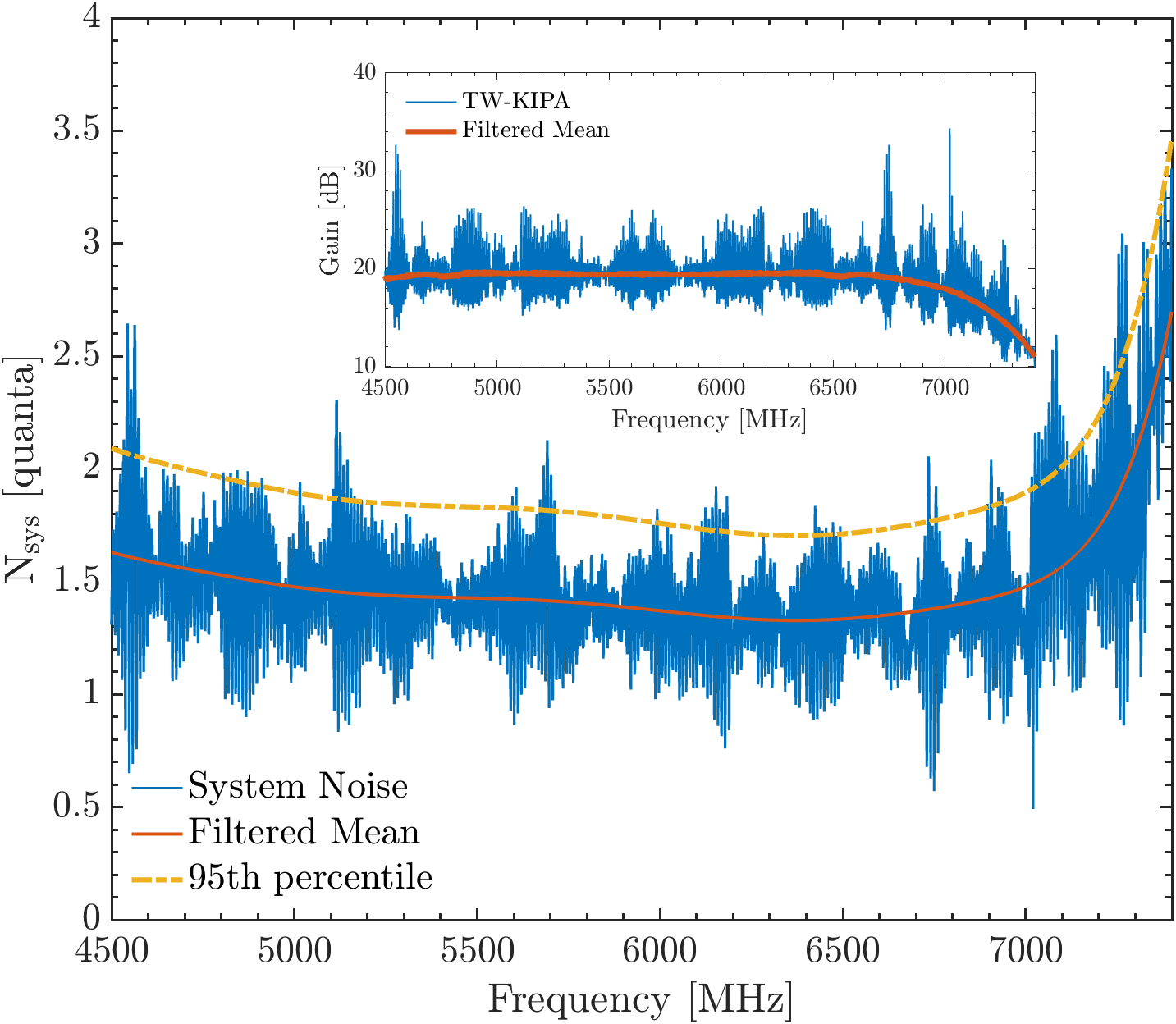}
    \caption{The overall system noise with 95th-percentile upper bound (yellow dashed), referred to the input of the parametric amplifier along with \textit{inset} gain of the TW-KIPA in the frequency regime of the experiment.}
    \label{fig:systemnoise}
\end{figure}

Our experimental procedure involves four spectrum analyzer measurements: the power emitted by the dish $P_\text{\rm d}$, the power emitted by the reference load $P_\text{\rm ref}$, and the power emitted by the hot and cold loads after the parametric amplifier ($P_{\rm H}$ and $P_{\rm C}$). The latter three measurements combined with $G_{\rm pa}$ allow us to refer the spectrum analyzer noise level to the number of noise quanta at the input of the parametric amplifier ($N_{\rm pa}$) in a `Y-factor' style measurement as described in \cite{klimovich2022traveling}. The overall system noise is dominated by the HEMT and prior components and is approximated by $N_\text{\rm sys} = N_\text{\rm pa} + N_\text{\rm HEMT} / (G_\text{\rm pa} A)$. The attenuation $A$ between the parametric amplifier and HEMT is based on measured $S_{21}$ transmission reference data for individual components and is taken to be a conservative estimate of -4~dB in this analysis. The gain of the parametric amplifier was only recorded above 4.5 GHz so we use the mirror image of the corresponding idler gain above $f_p/2$ for the frequencies below it. In the 1.15 GHz range where the data is present, this estimate agrees to within a mean absolute deviation of 0.05~dB and is dominated by the excess in our estimate of $A$. The frequency dependent HEMT noise is extracted from its datasheet. The noise result across frequencies is shown in Figure \ref{fig:systemnoise}, with the more than 20~dB gain of the amplifier shown in the \textit{inset}. The large-scale structures visible in this result stem from the impedance linked ripples of both the TW-KIPA operation as well as in switching the cold switch between hot load, cold load, and amplifier channels. Since using the calculated value directly would occasionally result in the unphysical choice of $N_{\rm sys} < 1$, we instead choose a conservative 95th-percentile upper bound, outlined by the dashed line in the figure, of $N_{\rm sys} \approx 2.1$. 

\begin{figure*}[!htp]
    \centering
    \includegraphics[width=\textwidth]{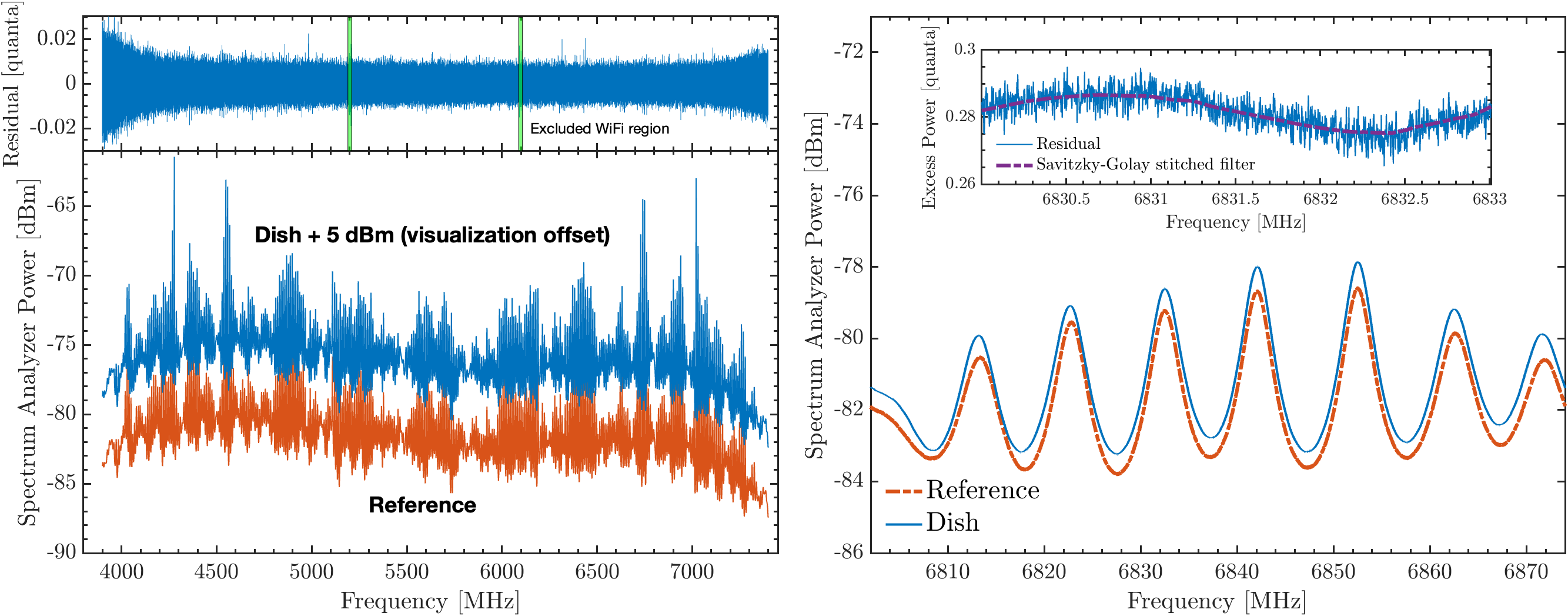}
    \caption{\textit{Bottom Left: } Integrated reference and dish data (offset by +5~dBm for visualization purposes) taken over the 8~hr. exposure. \textit{Right: } Zoom into the 6.84 GHz region, showing both the impedance mismatch ripple structure and excess broadband power present in the dish data. \textit{Right inset: } Residual of the dish and reference, showing the ripple mismatch between datasets, which introduces further sinusoidal structure. Dashed (purple) line is the Savitzky-Golay filter, as described in the text, used to compensate for these effects. \textit{Top Left: } The final filtered residual over the 3.5~GHz region of interest, with green band regions excluded from the analysis due to WiFi interference effects.}
    \label{fig:raw_data}
\end{figure*}

Next, we integrate the signal emitted by both the dish and reference load for 8.27 hours each. This sweep is performed from 3.9 to 7.4 GHz, covering the 3.5 GHz bandwidth centered around half the pump frequency ($\nu_{\rm p}=11.298$~GHz) of the parametric amplifier within which there is $>15$~dB of gain, and averaged over 1000 data runs, with resultant datasets as seen in Fig. \ref{fig:raw_data} \textit{bottom left}. The spectrum analyzer has an instantaneous bandwidth of approximately 160 MHz utilizing a high-speed digitizer followed by a digital FFT. The spectrum analyzer was configured to use a flat top window function with a resolution bandwidth of 3 kHz and video bandwidth of 3 Hz, providing for a bin size of $\Delta_{\rm b}=762$~Hz and a per bin integration time of approx. 5.5 mins. These choices reflect a balance between maximizing the SNR of the signal by matching the expected frequency dispersion of the dark matter ($\delta \nu \approx 10^{-6} \nu$, $\mathcal{O}$(kHz)) due to the relative velocity of the galactic dark matter halo, and the bandwidth limited readout speed of the analyzer.

Switching the cold switch at the input of our system between the attenuators and horn antenna also results in a shift of the ripple structure due to the altered path length for the standing wave reflections in the setup and the likely differential between the 377~$\Omega$ free-space impedance and 50~$\Omega$ RF components. Because these components were connected using a similar length of coaxial cable, the change is minor compared to the effect seen from altering the configuration of the cold switch used in the Y-factor measurement. We also see a broadband $\sim$0.4~dB excess in the dish spectrum, attributed to environmental emission picked up by the antenna. In conjunction, these differences create a frequency-dependent offset in the residual as seen in Fig. \ref{fig:raw_data} \textit{right} between the dish data and reference that must be modeled. We compensate for these effect by employing a method similar to that used in cavity experiments \cite{brubaker2017haystac} \textemdash\ applying a combination of third order Savitzky-Golay Filters with a 2.5 MHz (250 kHz) window to remove structures within the residual broader than 1 MHz (100 kHz). These two fits are smoothly stitched together using a weighted average with weights $(w_2 / w_1)^2$ and $1 - (w_2 / w_1)^2$ where $w_{1,2}$ is the sum of the residual over a 500 kHz Blackman-Harris window centered at each data point for the fit. The resulting residual, found to be Gaussian over 10~MHz windows, is seen in Figure \ref{fig:raw_data} \textit{top left} and shows that fluctuations in the measurement have been averaged to $\sim$0.01 quanta at the input of the parametric amplifier. We exclude two data regions in the dish dataset with large noise features, at 5.19\textendash5.21 GHz (and its idler), traced to the existence and use of WiFi channel 40 \cite{IEEEStandard} in the lab. Additionally, we sum the power of each neighboring set of five bins to effectively de-correlate the spectrum analyzer effect of using overlapping windows on the order of the flat-top width \cite{KbandDM1, brubaker2017haystac}.
 
The dark matter signal model we use is a literature standard isothermal Maxwellian velocity distribution with escape velocity $\rm{v_{esc}}$ = 544~km$\cdot$s$^{-1}$, mean $\rm{v_0}$ = 220~km$\cdot$s$^{-1}$, and periodic Earth motion with velocity $\rm{v_E}$ = 232~km$\cdot$s$^{-1}$. The velocity profile introduced frequency dispersion results in a relation between hidden photon frequency and dark matter velocity $v$ of $h\nu_{\rm HP} = m_{\rm DM}c^2/\sqrt{1-(v/c)^2}$.

\begin{figure}[!ht]
    \centering
    \includegraphics[width=0.45\textwidth]{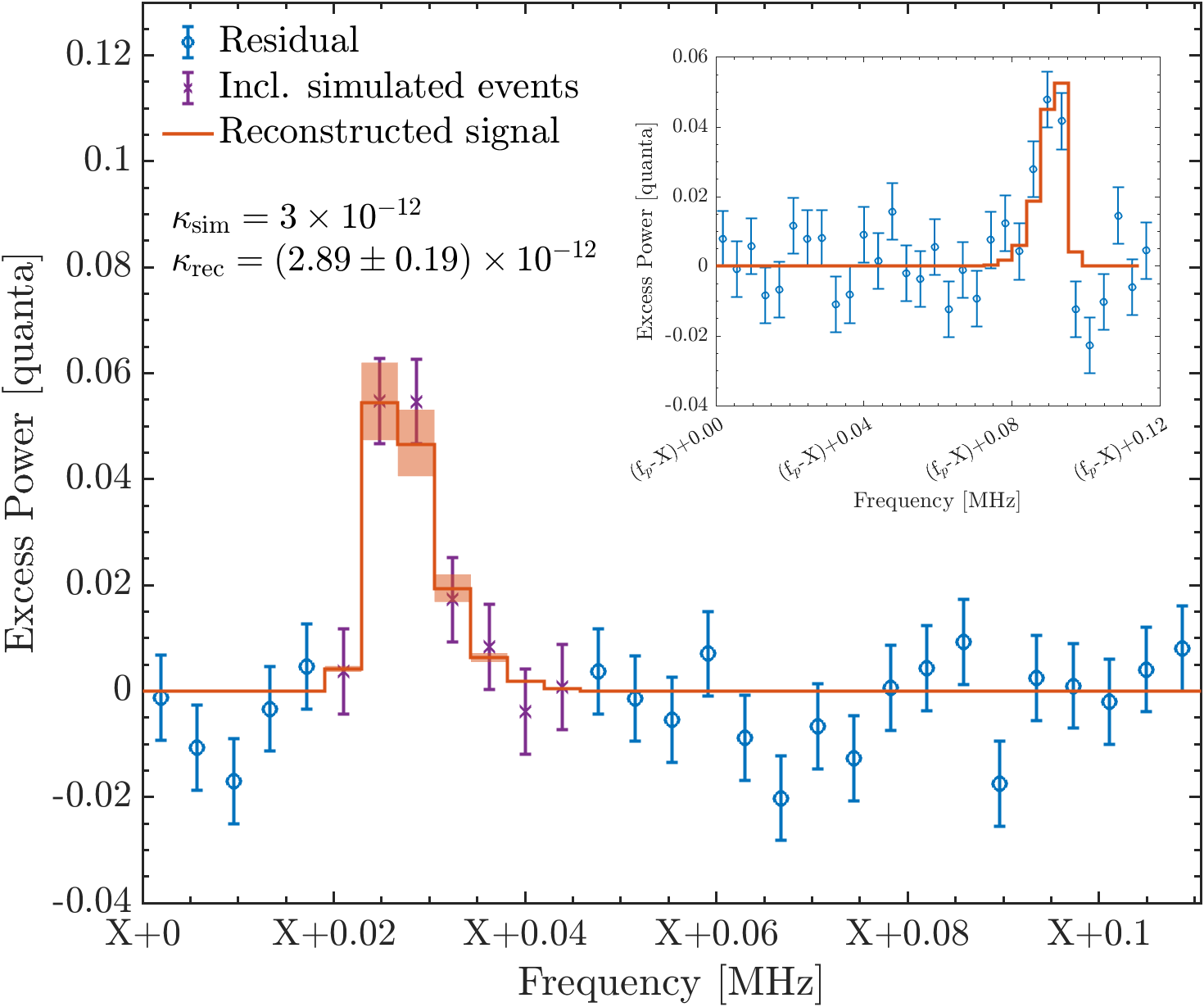}
    \caption{Example of a signal at \hpfreq=7~GHz with \km = 3$\times$\powert{10}{-12}, injected into a data window around 4~GHz, and overlaid with the signal reconstruction. \textit{Inset: } corresponding idler location and resultant fit.}
    \label{fig:fit}
\end{figure}

Constraining the signal power is then done via a two-step process. First, we implement a likelihood analysis in (\km, \mdm) space. We generate a signal and idler model for a given frequency \hpfreq, stepping through the data in $\delta$\hpfreq$\sim$10~kHz intervals, which are then convolved with an experimentally measured flat-top spectrum analyzer response to a single frequency tone. The fitting range is set at [\hpfreq\ - 25~kHz, \hpfreq\ + 75 kHz] and mirrored for the idler. An example of the expected spectral shape can be seen in Fig. \ref{fig:fit}, for a simulated signal injected into the raw dataset around 4~GHz \footnote{Chosen because no signal mode will be supported by the antenna at that frequency.}. Notably, a mirrored version of the signal with an amplitude scaling of $1 - 1/G_{\rm pa}$ should also appear at the parametric amplifier idler frequency as a result of the amplification process, see Fig. \ref{fig:fit} \textit{inset}. Because this ratio is near unity, any dark matter signal will effectively appear twice within the measurement, doubling the integration time in terms of its signal to noise ratio. The combined signal and idler model can then used to search for excesses as a function of \km\ within the data according to the expected power given by Equation \ref{eq:DMpower}. We minimize the binned Gaussian negative log-likelihood $\mathcal{LL}$ defined as,

\begin{align}
\begin{split}
    \mathcal{LL} = & \sum_{\rm Idler} \rm{Log}(\sigma_I^2) + \frac{1}{\sigma_I^2} \left(P_I^{\rm model} - P_I \right) \\
            + & \sum_{\rm Signal}  \rm{Log}(\sigma_S^2) + \frac{1}{\sigma_S^2} \left(P_S^{\rm model} - P_S \right)  \\
    \rm{with} \quad & \sigma_{S,I}^2 =\, \sigma_{\rm white}^2 + \sigma_{\rm em}^2 + N_{S,I} 
\end{split}
\end{align}
for per bin power $P_{S,I}$, model power $P^{\rm model}$ (derived from Eq. \ref{eq:DMpower}) and bin error terms $\sigma_{S,I}$. The latter is composed of a white noise term estimated from two neighboring non-overlapping fit windows, a thermal emission term accounting for the broadband subtraction between the dish and reference, and a bin dependent shot noise term $N_{i} \equiv b_{i} \cdot P_{i}^{\rm model} \tau / (\rm{h} \nu_{\rm HP})$, based on the expected signal weight $b_i$ and exposure time $\tau$, to account for potentially very weak signals from small \km. Again note that due to idler mirroring, any idler power is fully defined by the signal model $\Sigma P_I = (1-1/G_{\rm pa})\Sigma P_S$.

\begin{figure*}[!ht]
    \centering
    \includegraphics[width=1.\textwidth]{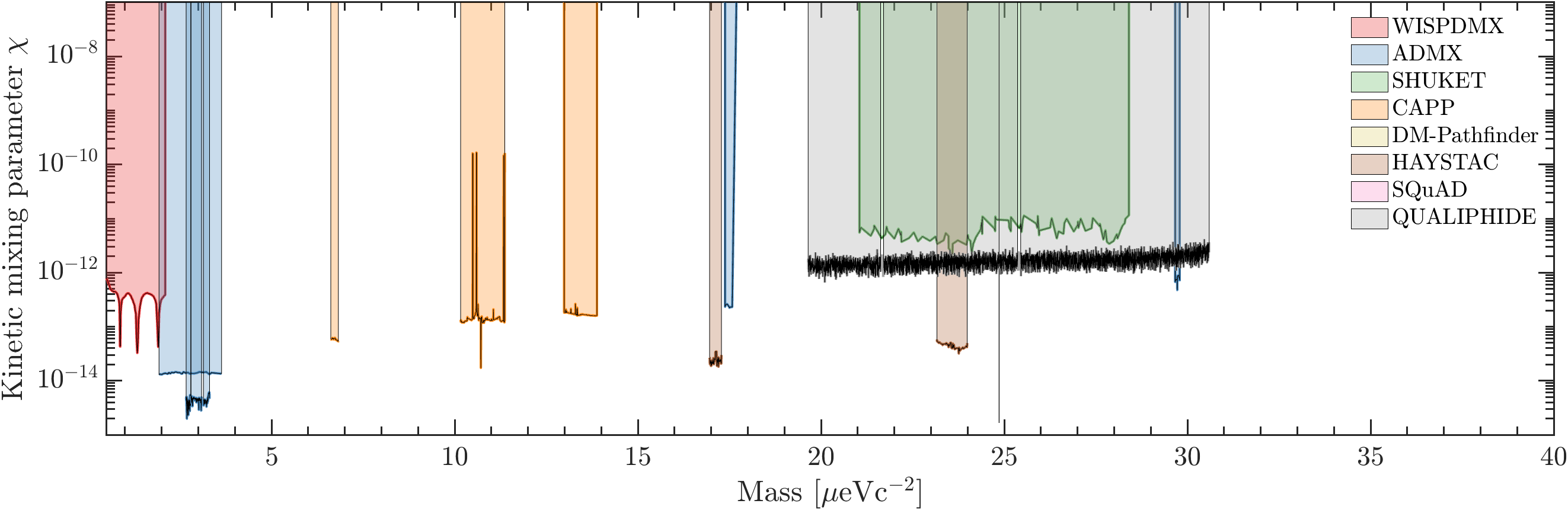}
    \caption{95\% confidence limits (gray) on the kinetic mixing parameter \km\ as obtained by QUALIPHIDE, with cut-out regions, overlaid with summarized limits from other experiments \cite{{AxionLimits, *caputo2021dark}} (see references for links to original experiment publications).}
    \label{fig:limits}
\end{figure*}

We find minima consistent with a non-zero signal expectation (i.e. local `$p$-values' of $\sim$10$^{-4}$) for certain mass values. However we account for the look-elsewhere effect with the methodology outlined in \cite{foster2018}, where we use a Monte Carlo simulation to determine the number of independent frequency windows ($\gtrapprox 10^4$) in the sample and re-evaluate the discovery significance. This analysis drops the significance of an excess in any bin below $\sim$1$\sigma$, implying we find no robust presence of hidden photon dark matter. 

Next, we compute the 95\% confidence limits by using the test-statistic $\Lambda = 2 \left(\mathcal{LL}_\text{model} - \mathcal{LL}_\text{min} \right)$ in conjunction with Wilks theorem. We verify that $\Lambda$ is $\chi^2_{\rm 1 dof}$ distributed as expected via another Monte Carlo simulation of 1000 injected and reconstructed signals each for a set of 10 randomly chosen \hpfreq\ spanning the search space. The resulting 95\% confidence limits on \km from QUALIPHIDE, compared with limits set by other experiments \cite{caputo2021dark}, are shown in Fig. \ref{fig:limits}. 

A few checks are performed to ensure robustness of the results. To ensure that the data cleaning process did not bias any reconstruction, we Monte Carlo inject 1000 signals of \km\ between 0.7\textendash4$\times$10$^{-12}$ and \hpfreq\ between 4\textendash8~GHz into a 100~MHz window around the 4~GHz region and verify accurate reconstruction of \km, an example of which is again shown by Fig. \ref{fig:fit}. We modify the Savitzky-Golay filter parameters (order$^{+2}_{-1}$ and frequencies $\pm$~10\%), which change \km\ limits at the 10\% level. Changes to the power-coupling due to effects like dish-antenna misalignment, systematics in the frequency dependent antenna gain, and diffraction effects, are modeled through changes in $\epsilon_c$ and are seen to follow the expected \km$\propto P_{\rm det}^2$ relation at the output of the analysis.

In summary, we have conducted a search for hidden photon dark matter and found no presence of a signal for most masses in the range between [1.97, 3.05]$\times10^{-5}$~\evc, excluding a kinetic mixing parameter $\km$ at the 95\% confidence level to just over \powert{10}{-12} across that frequency range. This result represents the first use of wideband quantum limited amplification to hunt for dark matter. The parasitic nature of this experiment, conducted in one day with an undersized dish relative to the cryostat volume, indicates that such quantum limited dish experiments can be a fruitful avenue for future searches. The forthcoming availability of K to W band TW-KIPAs \cite{che2017superconducting} and theoretically very well-motivated post-inflationary axion dark matter at $\sim$65~\muevc\ \cite{buschmann2022dark} (which can be looked for in a similar manner), provide for a natural next evolution of QUALIPHIDE.

We would like to thank Sunil Golwala for useful conversations relating to light dark matter searches. The research was carried out at the Jet Propulsion Laboratory, California Institute of Technology, under a contract with the National Aeronautics and Space Administration (80NM0018D0004). K.R. is supported by the Troesh family fellowship at the California Institute of Technology. 

\bibliography{refs.bib}

\end{document}